\documentclass[11pt]{article}

\usepackage[english]{babel}

\usepackage[affil-it]{authblk} 
\usepackage{etoolbox}
\usepackage{lmodern}
\DeclareUnicodeCharacter{2212}{-} 

\usepackage[letterpaper,top=2cm,bottom=2cm,left=3cm,right=3cm,marginparwidth=1.75cm]{geometry} 

\usepackage{amsmath}
\usepackage{graphicx}
\usepackage[colorlinks=true, allcolors=blue]{hyperref}

\title{Spin waves in Co$_{2}$FeGe films}

\author[1,2,3]{Popadiuk D.}
\author[4]{Vovk A.}
\author[4]{Bunyaev S.A.}
\author[4]{Kakazei G.N.}
\author[4]{Araujo J.P.}
\author[5]{Strichovanec P.}
\author[5]{Algarabel P.A.}
\author[2]{Golub V.}
\author[1]{Kravets A.}
\author[1]{Korenivski V.}
\author[3]{Trzaskowska A.}

\affil[1]{\footnotesize Nanostructure Physics, Royal Institute of Technology, 10691 Stockholm, Sweden}
\affil[2]{\footnotesize Institute of Magnetism of NAS of Ukraine and MES of Ukraine, 03142 Kyiv, Ukraine}
\affil[3]{\footnotesize Institute of Spintronics and Quantum Information, Faculty of Physics, Adam Mickiewicz University, Pozna\'n, Uniwersytetu Pozna\'nskiego 2, 61-614 Pozna\'n, Poland}
\affil[4]{\footnotesize Departamento de Física e Astronomia, Institute of Physics for Advanced Materials, Nanotechnology and Photonics, Universidade do Porto, 4169-007 Porto, Portugal}
\affil[5]{\footnotesize Instituto de Nanociencia y Materiales de Aragón, Universidad de Zaragoza—CSIC, Campus Río Ebro, 50018 Zaragoza, Spain}

\date{}

\begin{document}
\maketitle

\begin{abstract}
The dynamic magnetic properties of Full Heusler alloy thin films of Co$_2$FeGe, grown on MgO (001) substrates under different thermal conditions, were investigated. Brillouin light scattering and ferromagnetic resonance measurements revealed that depositing at room temperature followed by annealing at 300\textdegree C for 1 hour produces the best results for maximizing magnetization, exchange stiffness, and minimizing spin-dynamic dissipation in the films, which are desirable characteristics for high-speed spintronic devices. Additionally, strong hybridization of spin waves in the Damon-Eshbach geometry was observed, which is attractive for applications in magnonic signal processing circuits.
\end{abstract}

\section{Introduction}

Heusler alloys provide a wide range of potential applications, due to the tunability of their physical properties through modifications in chemical composition, atomic ordering, and microstructural changes~\cite{1Fesler_Book1_2016}. Full Heusler alloys of composition X$_2$YZ, such as Co$_2$MnX (X = Ge, Si, Sn)~\cite{Sharan2021, CANDAN2013215,ELKRIMI2020128707,ISHIDA19981}, Fe$_2$CoAl~\cite{Rai2020,AHMAD2021168449}, and Fe$_2$YAl (Y = Ni, Mn, Cr)~\cite{SHARMA2013142}, Co$_2$FeAl~\cite{Ahmad_2021}, have recently attracted significant interest from the research community. These materials with their half-metallicity~\cite{MOUATASSIME2021122534, SALEEM2022114947, LEI20115187}, giant magnetocaloric effect~\cite{Ahmad_2021}, and high thermoelastic stability~\cite{SALEEM2022114947} are promising for technological applications in magnetic refrigeration and spintronics. Heusler alloys can resemble binary semiconductors~\cite{KACIMI2014451} and exhibit a diverse array of properties depending on their chemical composition. The latter results in variations in the number of valence electrons that affect the chemical and physical properties of the materials~\cite{Pan2016,GRAF20111}. Heusler alloys can exhibit superconducting properties~\cite{Hoffmann2023, NALEVANKO2024108231} as well as semiconducting characteristics, making them excellent candidates for thermoelectric applications. Furthermore, Heusler alloys show typical properties of topological insulators~\cite{ZHILIN2020}, interesting magneto-optic properties~\cite{Kumar2022}, and antiferromagnetic characteristics~\cite{ALVARADOLEYVA2024171760}. Theoretical and experimental research is ongoing to discover new Heusler alloys aimed at specific applications. 

Full Heusler alloy (FHA) films share some of these desirable magnetic and electronic characteristics and are seen as promising candidates for various multifunctional applications. Previous studies~\cite{Ryabchenko_2013, Conca, Oogane_2015,nano14211745, TRZASKOWSKA201296} have shown that crystal structure, grain size, and surface roughness can be controlled by varying the composition and growth conditions of FHA films. As a result, both the static and dynamic magnetic properties of the films could be effectively tuned in a wide range. However, a standardized approach to heat treatment is yet to be established and, in practice, thermal processing is customized to each case, depending on the film's composition and specific preparation steps used.

Co$_2$FeGe and Co$_2$FeAl were recently predicted to possess half-metallicity with a gap of 1.2 eV and, consequently, 100\% spin-polarized electronic conduction~\cite{MOUATASSIME2021122534, Uvarov10.1063}. Microelectronic devices require thin films that need to be optimized with regard to their spin-electronic and spin-dynamic properties. In this work, we show how these properties are affected by the films' microstructure, which in turn is influenced by thermal processing. We find that a specific thermal regime can enhance the key spin-dynamic parameters important for applications. 

\section{Materials and methods}

\subsection{Samples}

Full Heusler alloy Co$_{2}$FeGe epitaxial films of 60 nm thickness were deposited onto $10 \times 10$~mm$^2$ MgO (001) substrates using magnetron co-sputtering (Orion-5 system by AJA International Inc.) at an argon pressure of 3~mTorr. The films were deposited at room temperature (sample RT) and 300\textdegree C (sample HT). Subsequently, a subset of the RT-deposited films was annealed at 300\textdegree C for 1 hour (sample RT-A). The elemental concentration was determined by EDX to be Co$_{48}$Fe$_{22}$Ge$_{30}$ to within ±1~at.~\%. More details on fabrication and characterization can be found in Refs.~\cite{nano11051229, nano14211745}.

\subsection{Measurements}

\begin{figure}[t]
\includegraphics[width=0.5\columnwidth]{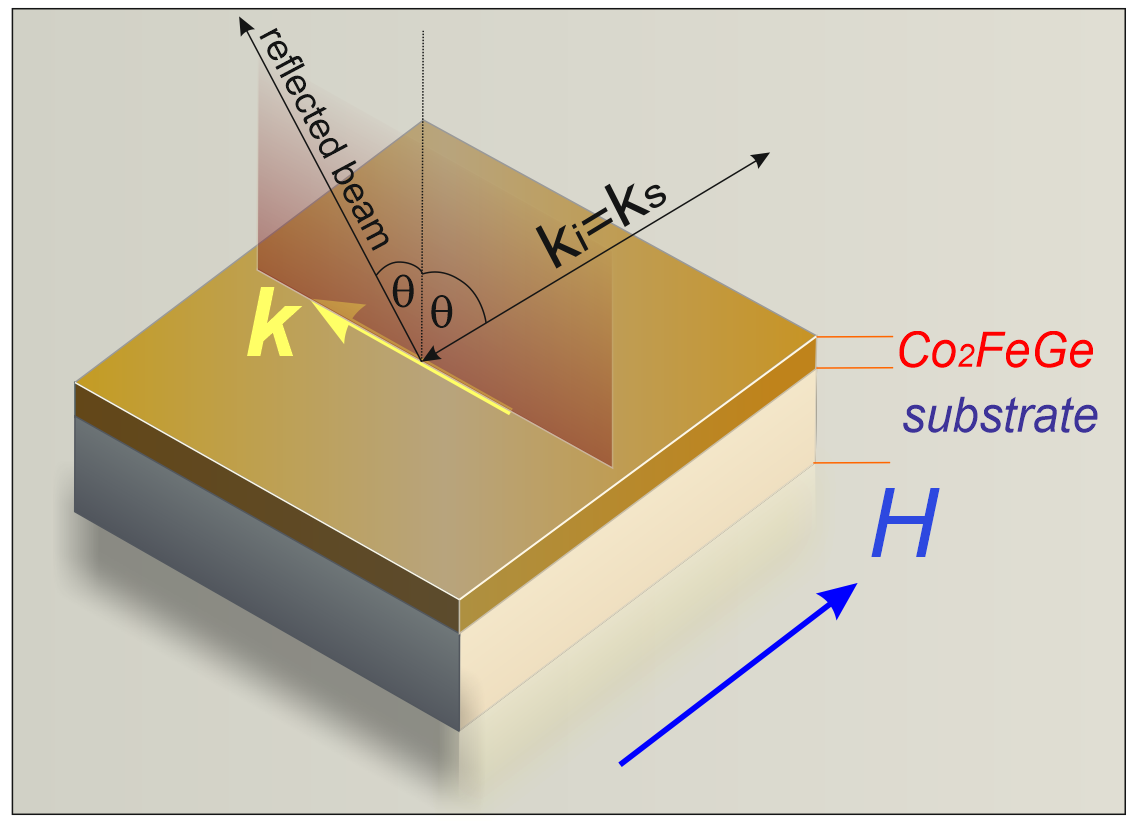}
\centering
\caption{\label{fig:DE} Damon-Eshbach geometry for BLS: $\mathbf{k}$ is wave vector of spin waves perpendicular to external field $H$ applied in sample's plane; $\mathbf{k}_i$– wave vector of incident light; $\mathbf{k}_s$ – wave vector of scattered light; $\theta$ is angle between incident light direction and sample's normal.}
\end{figure}

Brillouin Light Scattering (BLS) is an effective tool to characterize magnetization dynamics in Heusler alloys, recently used to study spin-waves in thick (120 nm) films of  Co$_{2}$Mn$_{0.6}$Fe$_{0.4}$Si~\cite{Sebastian_2015}. Here, measurements were performed at RT using a six-pass tandem Brillouin spectrometer (TFP2-HC, JRS)~\cite{TRZASKOWSKA201296, Scarponi_PhysRevX.7.031015, janardhanan2023investigation}. The light was generated by a frequency-stabilized DPSS laser operating at $\lambda_0$ = 532~nm (Coherent Verdi V5). Measurements were made in the backscattering geometry with the light polarization ps for magnons. The backscattered light was collected using f/8 optics with a focal length of 150 mm. The external DC field ($H$) was applied in the sample plane and perpendicular to the plane of light incidence, i.e., perpendicular to the wave vector \textit{\textbf{k}}. This geometry is called Damon-Eshbach or magnetostatic surface wave geometry; Fig.~\ref{fig:DE}.  Angle $\theta$ is the angle between the incident light and the normal to the sample plane, and the scattering wave vector \textit{\textbf{k}} was defined by the wavelength of the incident light and $\theta$ as $k =\frac{4\pi}{\lambda_0} \sin{\theta}$. The wave vector range was $0.01 \div 2.25\times10^{7}\ \mathrm{m}^{-1}$, with a resolution of $0.01 \times 10^{7}\ \mathrm{m}^{-1}$. Each spectrum was accumulated for 1 hour, and each peak in the spectrum was fitted using a Lorentzian curve. The frequency resolution in the BLS experiment was about 0.07 GHz.

Broadband ferromagnetic resonance (FMR) measurements were performed at RT utilizing a coplanar waveguide connected to a vector network analyzer Anritsu 37247 (Anritsu Corp.) operating in the frequency range 50~MHz -- 20~GHz. The external DC field ($H$) was applied in the plane of the films. The microwave field excitation ($\sim 0.1$~Oe) was applied in-plane and perpendicular to the DC field. The frequency spectra of the complex magnetic susceptibility $\chi(f)$ were subsequently derived from the raw S21 data using the practices outlined in references~\cite{Kalarickal,Tocac}. 

\section{Results and discussion}

\subsection{Structure}

The structure of the films under investigation was intensively studied by X-ray diffraction techniques in our previous work~\cite{nano14211745} . Out-of-plane (001) Co$_2$FeGe $||$ (001) MgO and in-plane [110] Co$_2$FeGe $||$ [100] MgO epitaxial growth conditions were confirmed. Deposition at the elevated temperature and heat treatment cause improvements of crystal quality and atomic ordering. For sample RT B2-type atomic ordering was observed, with Co atoms occupying their respective lattice positions while Fe and Ge were intermixed. Superlattice peaks with odd (hkl) indices (e.g., (111) and (113)), characteristic of fully atomically ordered  L2$_{1}$ phase, appeared in X-ray diffractograms for samples HT and RT-A. However, these reflections are also present for partially ordered DO$_3$, phase (Co and Fe or Co and Ge atoms are mixed at their crystallographic positions). Distinguishing between the L2$_{1}$ and DO$_3$ phases using conventional X-ray techniques is challenging due to the nearly identical atomic scattering factors of Co, Fe, and Ge atoms.~\cite{Balke01012008}. Although the formation of L2$_{1}$ atomic ordering is desirable due to its predicted half-metallicity~\cite{Ahmad_2021}, it is not always consistently achieved during fabrication~\cite{Guillemard_10.1063}. Sample RT exhibits a slight tetragonal distortion of Co$_2$FeGe cubic structure caused by in-plane tensile strain due to the lattice mismatch between Co$_2$FeGe and MgO. The diagonal of the MgO unit cell is ~0.5958 nm, while the bulk lattice parameter of Co$_2$FeGe is 0.5738 nm. As a result, the in-plane lattice parameter of the film (a = 0.5759 nm) was found to be larger, while the out-of-plane parameter (c = 0.5671 nm) was smaller compared to the bulk, with a c/a ratio of ~0.985. This distortion is reduced for samples HT and RT-A. The unit cell becomes almost cubic, manifesting the relaxation of the strain (c/a~0.999 for sample HT and ~0.99 for sample RT-A). Analysis of the X-ray rocking curve widths suggests that the epitaxial quality of the films improves on heating the substrate or annealing. Furthermore, X-ray reflectivity measurements show a reduction in surface roughness (from 1.4 nm for sample RT to ~0.9 nm for samples HT and RT-A) and an increase in film density (from 8.26 g/cm$^3$ to 8.3 g/cm$^3$, respectively). All of these factors point to the improvement of the microstructure of the films produced by heat treatment.

\subsection{Brillouin light scattering}

Detailed dispersion-field profiles of the spin waves (SW) in the studied films were measured using BLS spectroscopy. We performed two types of BLS measurements: SW frequency versus external magnetic field (1$\div$8~kOe; with films statically fully saturated), and SW frequency versus wave vector, in a fixed field of 1.6~kOe (for RT) and 1~kOe (for HT and RT-A). 

\begin{figure}[t]
\includegraphics[width=0.65\columnwidth]{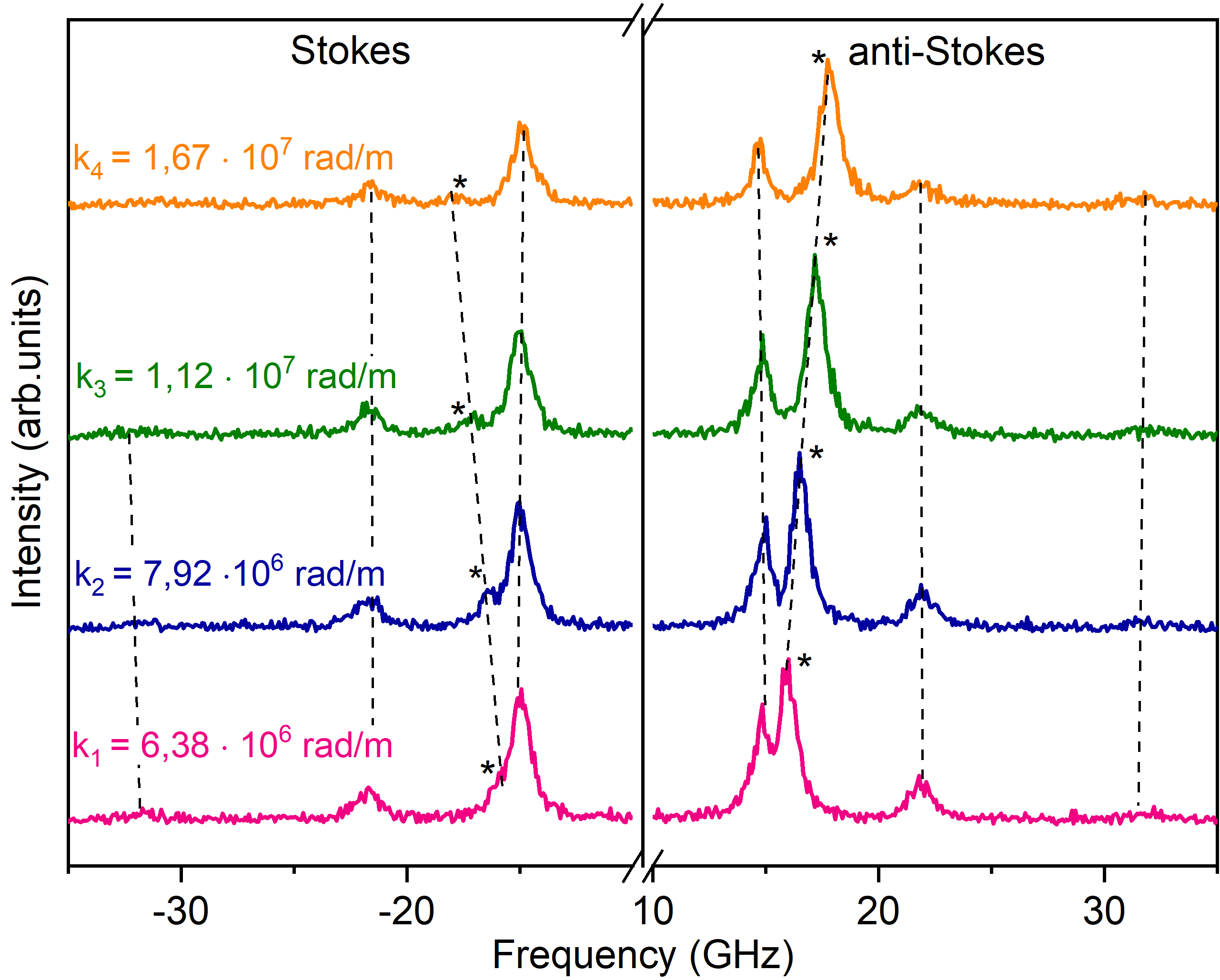}
\centering
\caption{\label{fig:BLS}BLS spectrum of sample RT measured in D-E geometry for different values of wave vector $\mathbf{k}$ in magnetic field of 1.6 kOe. Asterisks mark DE modes. Dashed lines are guides-to-the-eye indicating SW peak positions.}
\end{figure}

\begin{figure}
\includegraphics[width=0.55\columnwidth]{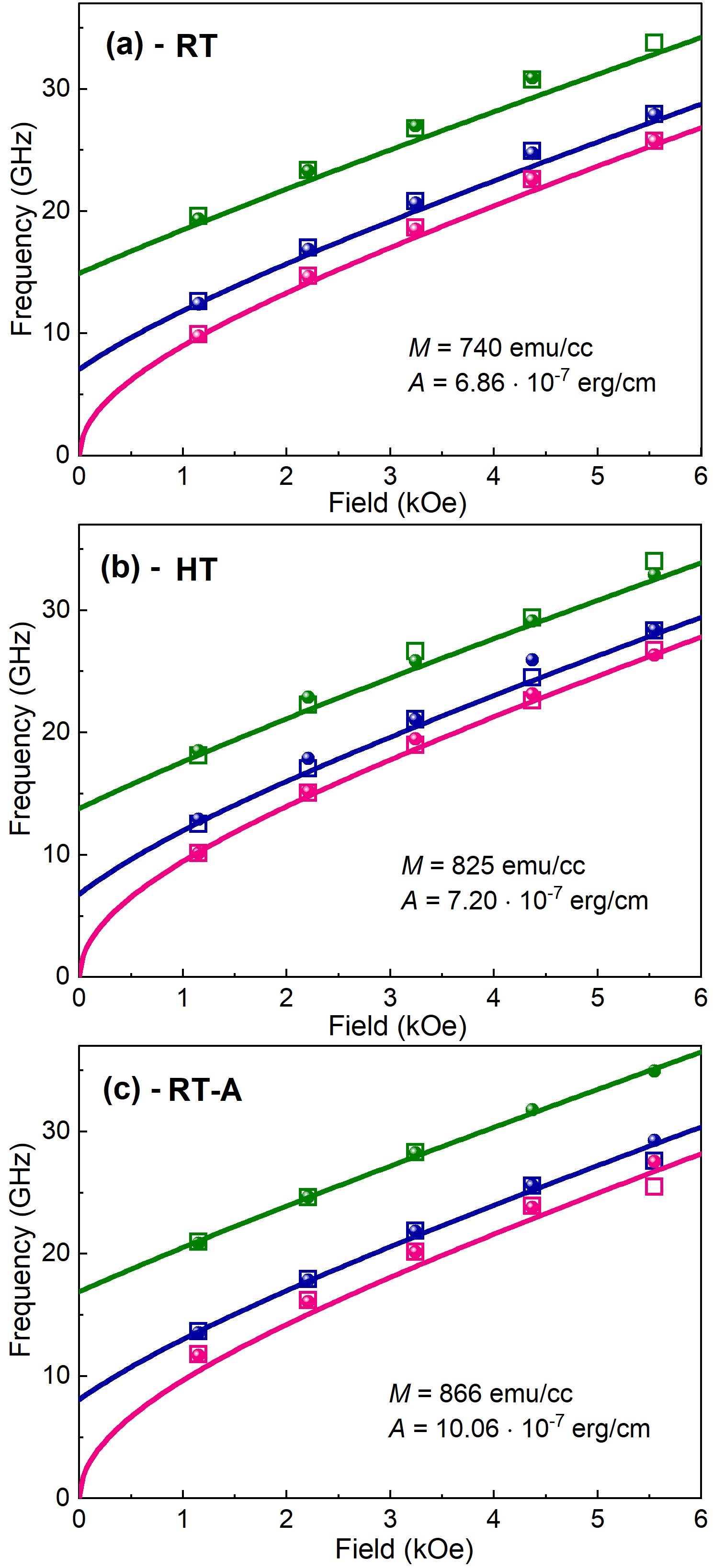}
\centering
\caption{\label{fig:BLS-H}BLS resonance frequencies for 3 SW modes vs. magnetic field for samples: (a) RT; (b) HT; (c) RT-A. Measurements were conducted in DE geometry for wave vector $\mathbf{k}$ close to zero; data points are shown as circles (Stokes) and squares (anti-Stokes). Solid lines are Kalinikos-Slavin model fits to DE, PSSW1, and PSSW2 data, with extracted effective magnetization and exchange stiffness parameters given in respective panels.}
\end{figure}

The raw BLS spectra, illustrated in Fig.~\ref{fig:BLS}, contain the Stokes (negative frequencies) and anti-Stokes (positive frequencies) spectral parts, which are asymmetric in intensity, expected under the nonreciprocity of the propagation of the Damon-Eshbach (DE) mode and the absence of symmetry of the associated magnon reflections~\cite{Kostylev,Sandercock1982}. The shift indicates the magnonic nature of the observed peaks~\cite{Sandercock1982}. To verify that the observed modes in the Brillouin spectra are spin waves, we measured their frequencies as a function of the magnetic field. For both the Stokes and anti-Stokes regions of the spectrum, the frequencies of the observed modes shift to higher values with increasing magnetic field strength, as shown in Fig.~\ref{fig:BLS-H}. The dependence on magnetic field confirms the magnonic nature of the observed modes (frequency of phononic modes would be unaffected by magnetic field). The modes in the measured BLS spectra correspond to the DE mode and perpendicular standing spin wave (PSSW) modes. 

\begin{figure}
\includegraphics[width=0.55\columnwidth]{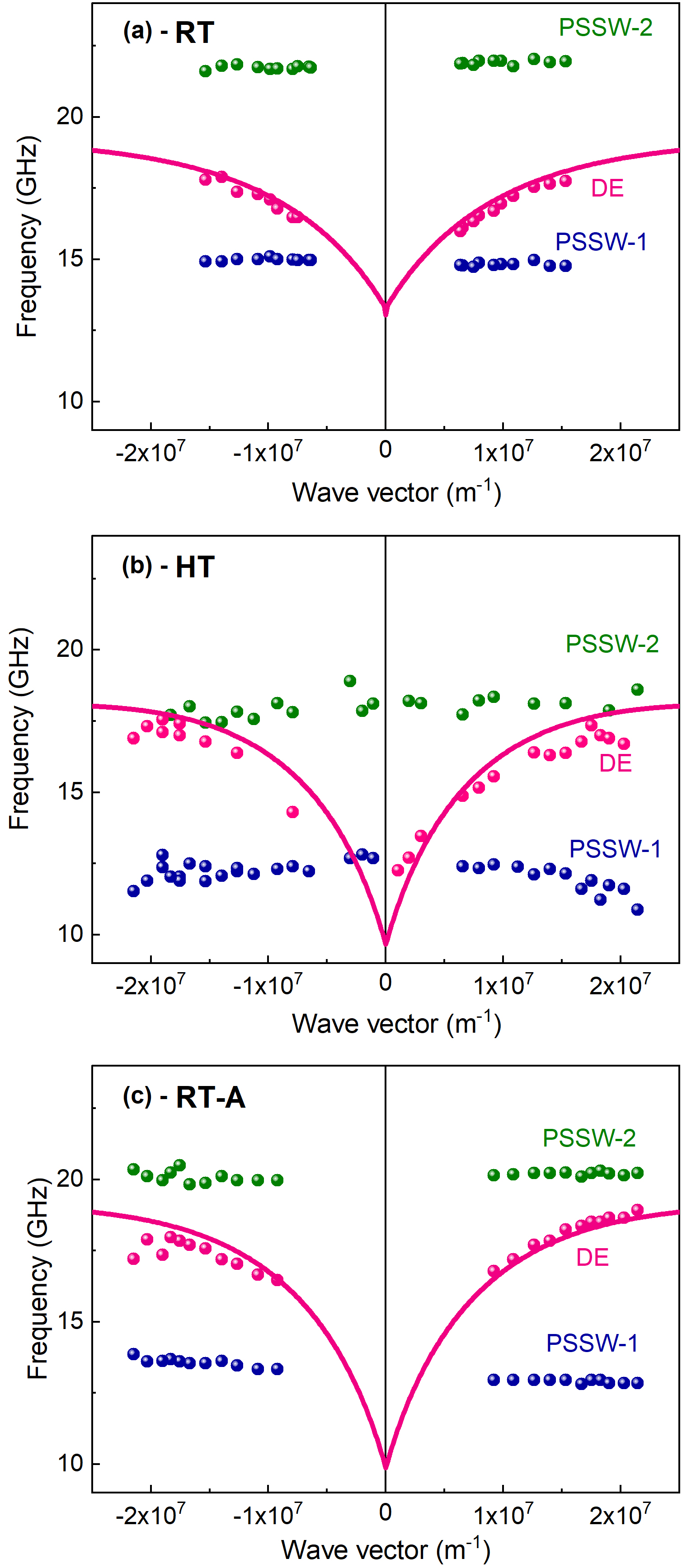}
\centering
\caption{\label{fig:BLS-f}BLS measured SW dispersion $\omega (k)$ for samples: (a) RT; (b) HT; (c) RT-A. DE mode is fitted with  Kalinikos-Slavin theory using same parameter set as for $\omega (H)$ of Fig.~\ref{fig:BLS-H}.} 
\end{figure}

The BLS dispersion data $\omega(k)$ for the different SW modes are shown in Fig.~\ref{fig:BLS-f}. Both parts of the BLS spectra contain the Damon-Eshbach surface mode, labeled with asterisks in Fig.~\ref{fig:BLS} and DE in Fig.~\ref{fig:BLS-f}, whose frequency increases with increasing wave vector. In contrast, the spectral positions of the PSSW modes (blue and green; no asterisks in Fig.~\ref{fig:BLS}) remain nearly constant~\cite{DEMOKRITOV}. This is in good agreement with previous studies~\cite{Ikeda2010}. The higher order dispersion curves fluctuate about fixed values in $\omega$ vs. $k$, which identifies them as perpendicular standing spin waves (labeled PSSW-1,2 in Fig.~\ref{fig:BLS-f}; even higher order modes are present, albeit barely above the noise floor).

Fitting the combined set of the $\omega (H,k)$ data shown in Figs.~\ref{fig:BLS-H},\ref{fig:BLS-f} using the Kalinikos-Slavin formalism~\cite{Kalinikos_1986} yields higher values for the effective magnetization $M$ and the exchange stiffness constant $A$ for samples HT and RT-A compared to sample RT~\footnote{We previously determined using cavity FMR measurements~\cite{nano14211745} that the samples under study possess fourth-order in-plane magnetocrystalline anisotropy: $H_a$(RT) = -55 Oe, $H_a$(HT)=27~Oe, and $H_a$(RT-A)=-35~Oe; these values were taken into account in fitting the FMR-BLS data}, with the values shown in the respective panels of Fig.~\ref{fig:BLS-H}. This indicates that deposition on a heated substrate or deposition at room temperature with subsequent annealing improves the crystal structure and, thereby, increases the degree of atomic and magnetic ordering in the material. Specifically, the obtained significantly enhanced $M$ and $A$ for sample RT-A associate with the observed higher degree of structural ordering in the L2$_1$ phase of the material (see Structure), known to have highly spin-polarized conductivity (approaching semi-metallic), which is attractive for spintronic devices such as sensors and oscillators.
 
The free-surface-spins case of the Kalinikos-Slavin theory~\cite{Kalinikos_1986} is found to fit our BLS data somewhat better than the pinned-surface-spins case (both cases symmetric regarding  top vs. bottom), although both yield good fits to the experimental data for $\omega (H)$ of Fig.~\ref{fig:BLS-H} as well as the DE mode in $\omega (k)$ of Fig.~\ref{fig:BLS-f}. The corresponding fits for the PSSW modes in $\omega(k)$ were not satisfactory, which is attributed to likely issues with crossing  between different SW-modes~\cite{Bondarenko-Kakazei}. Analysis of such cases is non-trivial and goes beyond the scope of this paper. The overall quality of the fitting procedure used, as it relates to extracting reliable material parameters, is nevertheless judged to be good, and is verified with a complementary FMR technique (see FMR section below). 

The dispersion relationships $\omega(k)$, readily obtained using BLS spectroscopy, are important characteristics of SW dynamics in the material. For our samples (Fig.~\ref{fig:BLS-f}) these consist of the DE mode as well as several PSSW modes. The frequency of the DE mode increases with $k$ for all samples. The PSSW modes are nearly constant with $k$ (slightly parabolic in form, clear on vertical scale magnification; seen in, e.g., PSSW-1,2 in (b)), which is an indication of a difference in intensity of absorption vs. emission of magnons in the system. Our modeling using the standard approach~\cite{Kalinikos_1986} yields for sample RT one hybridization point, where the DE and PSSW-1 modes cross, at $k_1 = \pm 6.6\times10^6~\mathrm{rad/m}$. For sample HT, a two-fold hybridization of the SW modes is observed for DE \& PSSW-1, PSSW-2, at $k_1 = \pm 4\times 10^6~\mathrm{rad/m}$, $k_2 = \pm 1.8\times10^6~\mathrm{rad/m}$, respectively. For sample RT-A, DE \& PSSW-1 hybridize at $k \approx \pm 3 \times 10^6~\mathrm{rad/m}$. In the region of the avoided crossing between the DE and PSSW dispersion branches, the group velocity of the spin waves decreases significantly because of the nearly flat dispersion. This phenomenon enables the localized "trapping" of spin waves, which can be utilized in magnonic applications for precise control of spin-wave propagation and the creation of magnonic waveguides. The strong coupling and hybridization of the modes enable coherent information exchange between the DE and PSSW modes, which is essential for efficient data processing in magnonic logic and memory applications~\cite{PhysRevMaterials.5.064411, dion2024ultrastrong, PhysRevApplied.16.054033}. Key material parameter for these is low spin damping -- minimal energy dissipation from spin wave excitations into lattice phonons.

\begin{figure}[t]
\includegraphics[width=0.5\columnwidth]{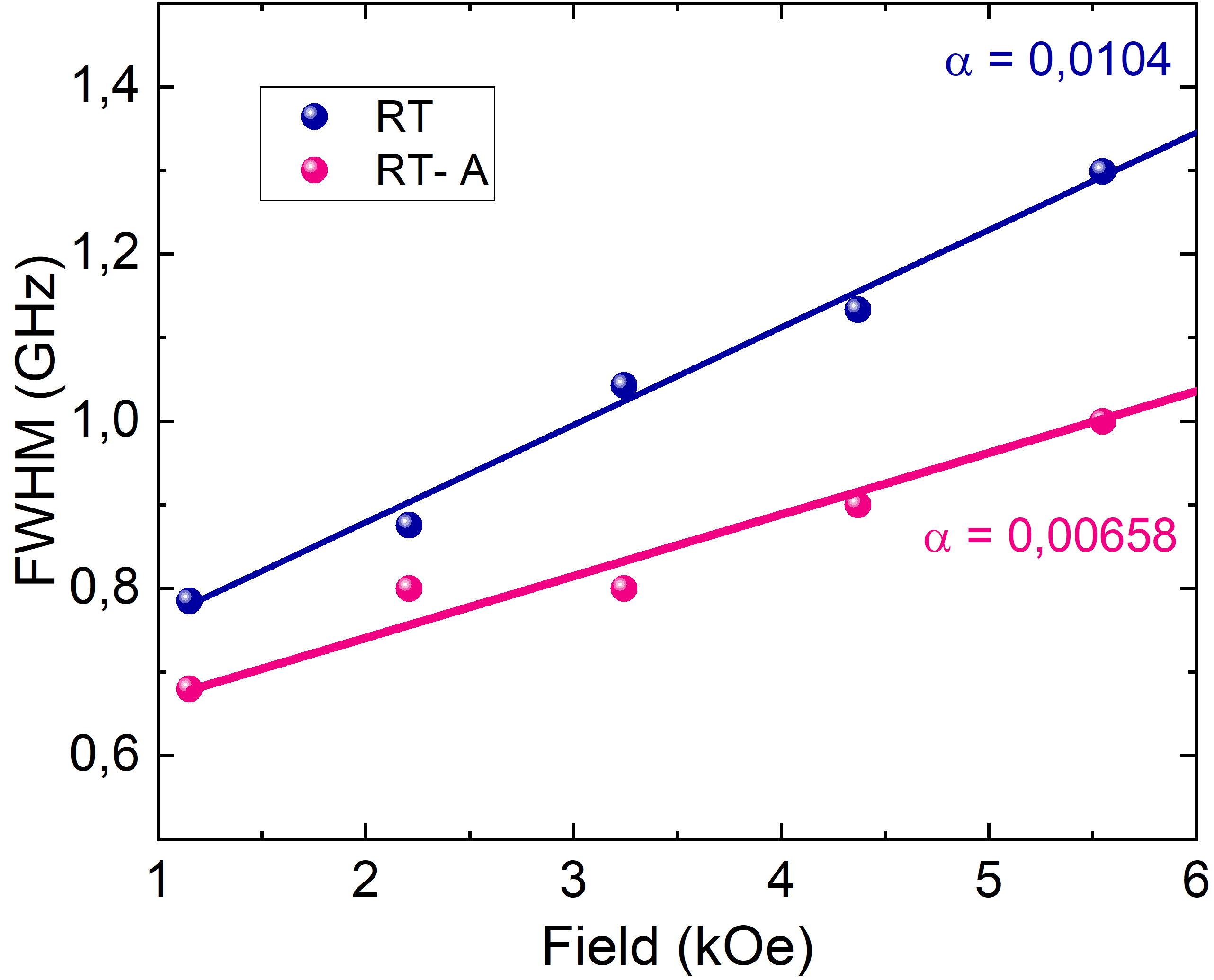}
\centering
\caption{\label{fig:BLS-FWHM}FWHM from BLS intensity-frequency spectra, $I(\omega)$, versus magnetic field for samples RT and RT-A. Solids lines show fits using Eq.~\ref{ec:equation3} yielding spin damping parameter $\alpha$.}
\end{figure}

BLS spectral linewidths are determined as the full width at half maximum (FWHM) in BLS signal intensity vs. field data, $I(\omega)$ vs. $H$, shown in Fig.~\ref{fig:BLS-FWHM}, and characterize the magnetization dissipation (spin damping) in the material. Although extracting spin damping values from BLS data is not common, it has been used in previous studies and showed good consistency with other techniques~\cite{PhysRevLett.120.157204,10.1063/1.4945685}, making it a viable tool in this context. Our BLS data for FWHM in Fig.~\ref{fig:BLS-FWHM} show a linear relationship with the applied magnetic field, which can be modeled by the following equation: 
      \begin{equation} \label{ec:equation3}
            \mathrm{FWHM}= \frac{2\alpha \gamma}{ \pi } H + \delta f_0,\\
     \end{equation}
where $\alpha$ represents the intrinsic spin damping (Gilbert damping parameter), and $\delta f_0$ accounts for extrinsic contributions to the linewidth, unrelated to the magnetic field, primarily due to instrumentational limitations and sample inhomogeneity. The obtained $\alpha$ values for as deposited (RT) and annealed (RT-A) samples are shown in Fig.~\ref{fig:BLS-FWHM} and give additional evidence for improved material quality (lower damping) after thermal treatment. Linear fitting yielded the correlation coefficients of 0.99 (RT) and 0.97 (RT-A), indicating high fit accuracy (no reliable fit could be obtained for sample HT, however). Thus, a significant reduction in the magnetization dissipation is clearly observed on annealing, which is another evidence of improved microstructure. The $\alpha$ values obtained and the trend on annealing obtained using BLS are in agreement with our recent measurements using broadband-FMR~\cite{nano14211745}: in excellent agreement for sample RT (0.01 vs. 0.089) and satisfactory agreement for sample RT-A (0.0066 vs. 0.0042); BLS had no satisfactory fit for sample HT whereas FMR yielded $\alpha =0.068$. 

The lowest spin damping found for sample RT-A combined with the highest magnetization and strongest exchange determined from $\omega (H,k)$ correlate well with the highest degree of atomic ordering and best microstructure observed after annealing. Low damping is key for SW propagation and thus for magnonic devices, where the $M$ and $A$ values should be known for deciding the device geometry and are not particularly restricted (larger $M$ improves inductive excitation and readout). Low damping is also important for spintronic devices such as GHz oscillators, combined with high spin-polarization (ideally half-metallicity), which our annealed material is expected to possess in its highly-ordered high-$M$, high-$A$ L2$_1$ phase.  

\subsection{Ferromagnetic resonance}

\begin{figure}[t]
\includegraphics[width=0.5\columnwidth]{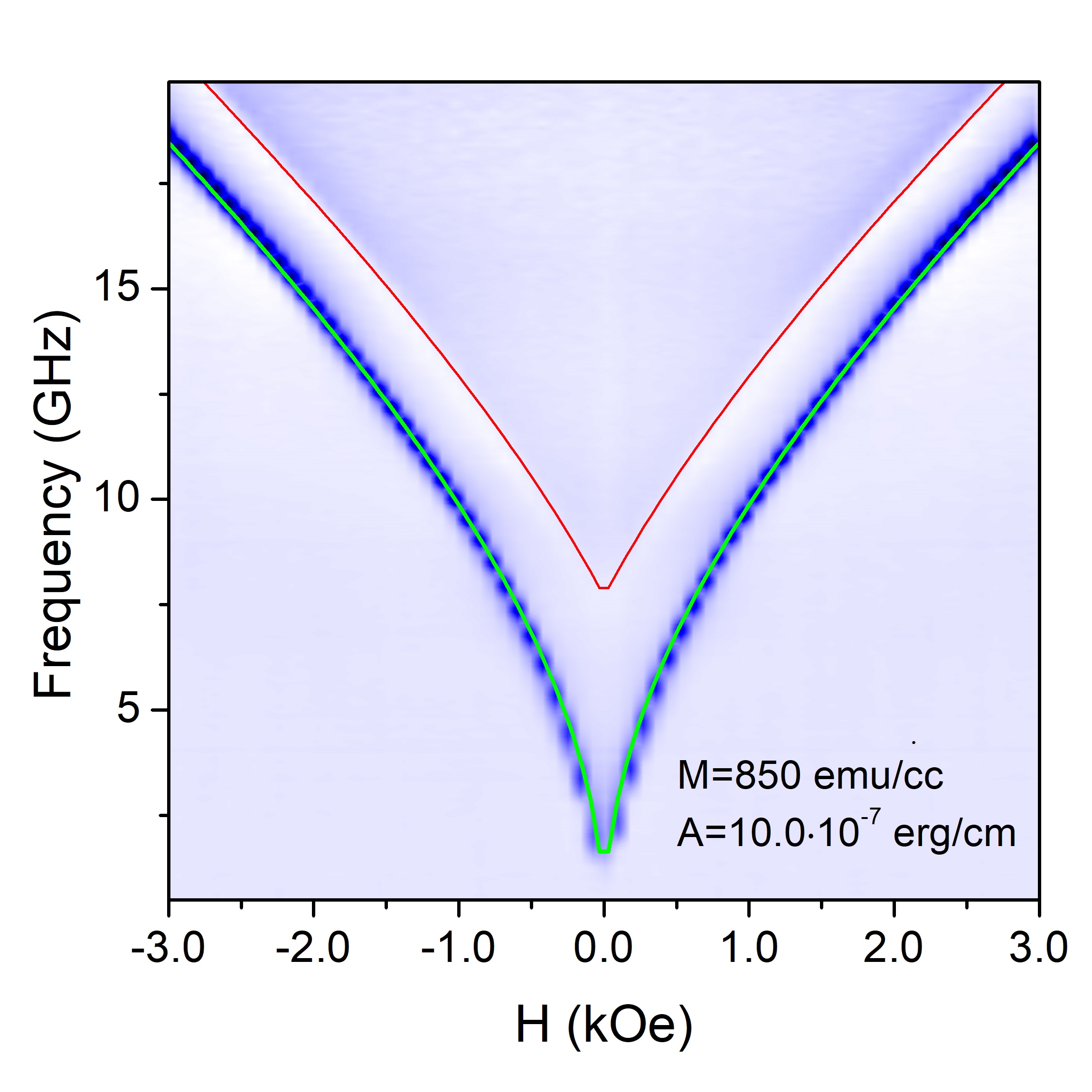}
\centering
\caption{\label{fig:FMR}FMR frequency-field map for sample RT-A used to extract effective magnetization and exchange-stiffness parameters. Solid lines show model fits to experimental data for uniform (green) and first PSSW (red) modes, with extracted parameters $M,A$ indicated.}
\end{figure}

In order to independently verify the BLS results, broadband FMR spectra were measured for the studied films as frequency sweeps in a fixed DC field ($H$) , from which the resonance frequencies can be determined. Figure~\ref{fig:FMR} illustrates the measured FMR $f$ vs. $H$ maps for sample RT-A, where the blue scale encodes the intensity of the FMR signal. Clearly seen is the resonance having Kittel's functional form for uniform FMR in thin films, as well as a trace of the first PSSW mode found at higher frequencies. The data for the other two samples differ only quantitatively. 

The FMR data were analyzed using the method outlined in Ref.~\cite{nano11051229}. The solid lines in Fig.~\ref{fig:FMR} represent the fits for the uniform mode (green) and PSSW-1 (red). The effective magnetization $M$ and exchange stiffness constant $A$ determined from the fits are shown in Fig.~\ref{fig:FMR}. For samples RT and HT they were 740 emu/cc, $6.6 \times 10^{-7}$ erg/cm and 830 emu/cc, $6.8 \times 10^{-7}$ erg/cm, respectively. The agreement between BLS and FMR is excellent, at most 2\% deviation for $M$ and $\leq$6\% for $A$. A lower exchange constant is associated with atomic disorder and dilution of magnetic atoms by non-magnetic~\cite{kittel1996solid}. Thus, the increase in $A$ for heat treated samples indirectly indicates the improved atomic ordering and microstructure of the films. All trends in the magnetic material parameters obtained using BLS are thus confirmed by FMR spectroscopy.

It may be informative to note that, from our broadband-FMR measurements, the inhomogeneous contribution to the resonance linewidth was estimated to be 62~Oe for the RT sample and 38~Oe for the RT-A sample. Such small values indicate that the films are essentially free from structural defects. Considering additionally their relatively large thickness, surface defects can be also neglected. Therefore, two-magnon scattering, caused predominantly by structural imperfections\cite{zakeri2007spin}, is unlikely to contribute to the spin-damping values observed for the studied films.

\section{Conclusions}

We have investigated the spin dynamic properties of Co$_2$FeGe films subject to different thermal processing during- and post-deposition. Analysis of the BLS and FMR measurements yields the effective magnetization, exchange stiffness, and SW damping in the films. We observe a clear trend toward higher magnetization and stronger exchange on heat treating the samples, with the highest values found for the sample deposited at room temperature and annealed for 1 hour (RT-A) at 300\textdegree C. This coincides with the improved atomic ordering and microstructure of the film. Maximizing these key magnetic parameters, with a concomitant reduction in spin damping, improves the material's magneto-dynamic, magneto-caloric, spin-transport, and other properties important for applications in spintronics and magnonics.
 
\section*{Acknowledgements}
Support from the Swedish Research Council (VR 2018-03526), Olle Engkvist Foundation (2020-207-0460), Wenner-Gren Foundation (GFU2022-0011), Swedish Strategic Research Council (SSF UKR24-0002) are gratefully acknowledged. The Portuguese team acknowledges the support through FCT – Portuguese Foundation for Science and Technology under the projects LA/P/0095/2020 (LaPMET), UIDB/04968/2020, UIDP/04968/2020, 2022.03564.PTDC (DrivenPhonon4Me), and SFRH/BPD/84948/2012 (A. Vovk). This work was partially supported by Spanish Ministerio de Economía y Competitividad through project PID2020-112914RB-I00, and from regional Gobierno de Aragón through project E28 20R including FEDER funding. The research leading to these results has received funding from the Norwegian Financial Mechanism 2014-2021 Project No. UMO-2020/37/K/ST3/02450 and from the National Science Centre of Poland, Grant No. UMO-2020/37/B/ ST3/03936.

\bibliographystyle{unsrt}
\bibliography{main}

\end{document}